\documentclass[12pt]{article}

\usepackage{epsfig}

\begin{document}

\title{Ball lightning as a possible manifestation of high-temperature
superconductivity in Nature}
\author{B.L. Birbrair\\
Petersburg Nuclear Physics Institute\\
Gatchina, 188300 St. Petersburg, Russia;\\
\em birbrair@thd.pnpi.spb.ru}
\date{}
\maketitle

\begin{abstract}
In the superconducting medium the circular current supported by  its
own magnetic field can exist giving rise to the possible underlying
mechanism for the ball lightning.
\end{abstract}

The example of such self-supporting object is provided by
superconducting circular current around the tube of the torus as shown
in Fig.1. The current consists of charged particles moving in a circle
of radius $r$. It is worth mentioning that such a motion is
two-dimensional the necessary condition for the superconductivity
\cite{1} thus being satisfied. The centrifugal force is balanced by
the Lorentz one so
\begin{equation}
\frac{m\gamma v^2}r=\mu\frac ec vH\ , \quad \gamma=(1-\beta^2)^{-1/2},
\quad \beta=\frac vc\ ,
\end{equation}
where $\mu$ is the magnetic permeability of the medium, $m$ is the rest
mass of the particle, $e$ is its charge, and $\gamma$ is the Lorentz
factor. It must be taken into account because the ball lightning is
luminous object the particle velocity $v$ thus being close to the light
one $c$ (the radiation is synchrotronic in the case under
consideration). The magnetic field within the tube of the torus is
\cite{2}
\begin{equation}
H\ =\ \frac{2I}{cR}\ .
\end{equation}
It is practically homogeneous when $r\ll R$ thus ensuring the circular
motion of the particles. The current strength $I$ is
\begin{equation}
I\ =\ \frac{Qv}{2\pi r}\ =\ \frac{ev}{2\pi r}\,N\ , \quad Q\ =\ Ne\ .
\end{equation}
$Q$ and $N$ being the total moving charge and the number of charged
particles respectively.

Putting Eqs. (2) and (3) into Eq.(1) we obtain the following connection
between the particle number $N$ and the radius of the torus
\begin{equation}
N\ =\ \frac{\pi mc^2\gamma}{\mu\,e^2}\ R
\end{equation}
and the following expression for the magnetic field
\begin{equation}
H\ =\ \frac{mc^2\gamma\beta}{\mu er}\ .
\end{equation}
The energy of the object under consideration is the sum of the magnetic
field one
\begin{equation}
{\cal E}_m\ =\ 2\pi^2r^2R\ \frac\mu{8\pi}\ H^2=\ \frac{\pi(\gamma-1)
(\gamma+1)(mc^2)^2}{4\mu\,e^2}\ R\ ,
\end{equation}
and the kinetic energy of the moving charges
\begin{equation}
{\cal E}_k\ =\ (\gamma-1)mc^2N\ =\
\frac{\pi\gamma(\gamma-1)(mc^2)^2}{\mu\,e^2}\ R
\end{equation}
the total energy thus being
\begin{equation}
{\cal E} \ =\ {\cal E}_m+{\cal E}_ k\ =\
\frac{\pi(\gamma-1)(5\gamma+1)(mc^2)^2}{4\mu\,e^2}\ R\ .
\end{equation}
To get the expression (6) we used the relation
$\gamma^2\beta^2=\gamma^2-1$, see Eq.(1).

The Lorentz factor $\gamma$ can be determined from the observed angular
frequency of the synchrotron radiation \cite{2}
\begin{equation}
\omega\ =\ \frac{2\pi c}\lambda\ =\ \frac{eH}{mc}\ \gamma^2\ =\
\frac{\beta\gamma^3c}{\mu r}\ =\
\frac{\gamma^2(\gamma^2-1)^{1/2}c}{\mu\,r}\ .
\end{equation}
In this way we get
\begin{equation}
\gamma^2(\gamma^2-1)^{1/2}\ =\ \frac{2\pi r}\lambda\ \mu\ ,
\end{equation}
where $\lambda$ is the wavelength of the radiation. The intensity of
the radiation is \cite{2}
\begin{equation}
S\ =\ \frac{2ce^4(\gamma^2-1)}{3(mc^2)^2}\ H^2\ =\
\frac{2ce^2(\gamma^2-1)^2}{3\mu^2r^2}\ .
\end{equation}

The calculations are performed assuming the charged particles to be
electrons and putting $\mu=1$. The average observed diameter of the ball
lightning (hereafter BL's) is 24~cm \cite{3,4}, but twice as large
diameters are also observed rather often \cite{4,7}. For this reason
the results are obtained for both the $R=12$~cm and $R=24$~cm values of
the torus radius. The tube radii are quite arbitrarily chosen as
$r=(1,2,3)$cm. It is worth mentioning in this connection that the
sphere is not the only observed form of the BL's:
many different forms including the torus are also observed \cite{5}.
The intervals of the Lorentz factor values are determined for the
visible light region running from $\lambda=7\cdot10^{-5}$cm (red) to
$\lambda=3.8\cdot10^{-5}$cm (violet) because the observed BL colours
cover all this region \cite{3}.

The results are shown in Table 1. Two features are important.
\begin{table}[h]
\caption{Intervals for the Lorentz factors, energies, and radiation.
The left and right bounds of each interval refer to the red and violet
lights respectively.}
\begin{center}
\begin{tabular}{|c|c|c|c|c|} \hline

& & \multicolumn{2}{c|}{${\cal E} \cdot10^3$~J} & \\ \cline{3-4}

$r$, cm & $\gamma$ & $R=12$~cm & $R=24$~cm & $S\cdot10^{-9}$W \\
\hline &&&&\\

1 & $44.8\div54.9$ & $29\div43$ & $58\div86$ & $1.85\div4.18$\\

2 & $56.4\div69.2$ & $46\div69$ & $92\div138$ & $1.16\div2.64$\\

3 & $64.6\div79.2$ & $60\div91$ & $120\div182$ & $0.89\div2.01$\\
\hline
\end{tabular} \end{center}
\end{table}

a. The energies are rather large, being practically the same as the
average value of 100~kJ for the outdoors observations of BL's \cite{6}.

b. At the same time the intensity of the radiation is rather small.
Both these features are characteristic for the exploding BL's \cite{5}.

In this way we showed  that the object with the similar properties to
those of the exploding BL's can exist in the superconducting medium
(the superconductivity must be high-temperature since there is no
reasons to assume the temperatures of BL's to be low). We do not know
whether such a medium arises in the atmospheric processes leading to
the BL's (this problem is out of the scope of the present work), but
the above results suggest that the exploding BL's may be a possible
evidence of this phenomenon.

The author is greatly indebted to Professor O.I.~Sumbaev for the
permanent attention to this work and Drs. V.L.~Alexeev and A.I.~Egorov
for stimulating discussions.

\newpage

\bigskip
\begin{center}
\begin{figure}[h]
\centerline{\epsfxsize=11.5cm\epsfbox{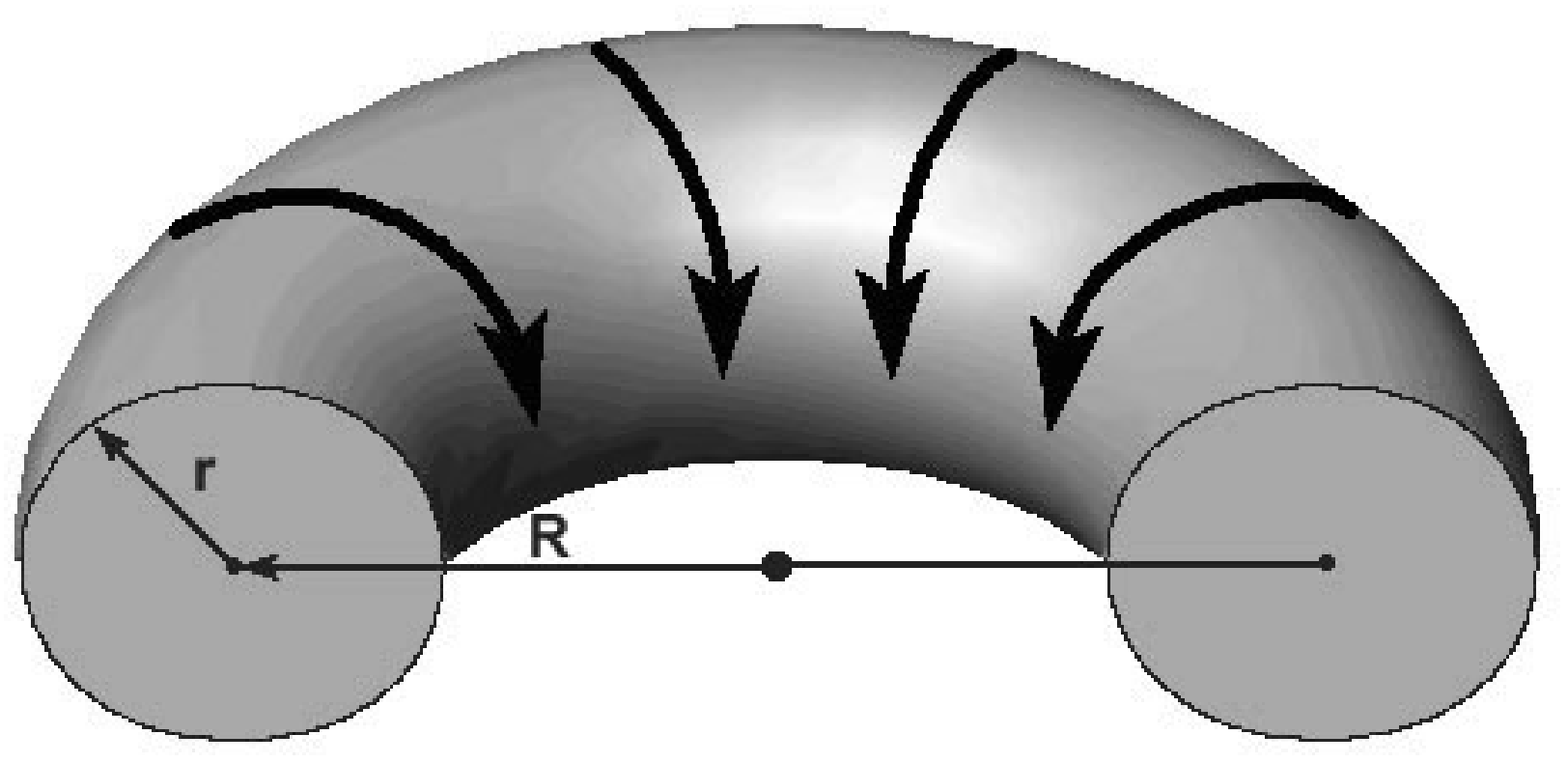}}
\caption{
The torus of radius $R$ and the tube radius $r$. The
superconducting surface current is shown by arrows.
}
\end{figure}
\end{center}

\end{document}